\documentclass[%
 reprint,
 amsmath,amssymb,
 aps,
]{revtex4-2}

\usepackage{graphicx}
\usepackage{dcolumn}
\usepackage{bm}
\usepackage{subfigure}
\usepackage{graphicx}
\usepackage[percent]{overpic}

\begin{document}


\title{Novel Light-Induced States in Triangular Metallic Magnet}

\author{Yao Wang}
\affiliation{School of Physcis and Optoelectronic Engineering, Shandong University of Technology, Shandong, 255000, China}
\email{yaowang@sdut.edu.cn}

\date{\today}

\begin{abstract}
Novel nonequilibrium states of magnet induced by light attract considerable attention both in nature of physics and apply.
In this work, we systematically explore the electronic and magnetic states of a double-exchange model on a triangular lattice under the irradiation of circularly polarized continuous wave field, by means of molecular dynamics calculation.
Several exotic nonequilibrium magnetic states are discovered, including a vortex state, long-range magnetic orders at the $\Gamma$ and $\textbf{K}/2$, as well as quasi(dynamical)-long-range magnetic order at the $\textbf{K}$ and $\textbf{M}$, respectively. Correspondingly, the evolution of electron bands and fillings are also uncovered. These results offer a promising candidate approach for the optical control of exotic magnetic and electronic states.
\end{abstract}

\maketitle


\section{\label{sec:intro}introduction}
Optical control of magnetism plays an increasingly important role in both fundamental and applied physics owing to its high speed, ease of handling, tunability of strength and frequency, extra phase and the polarization degree of freedom, and so on. Accessing all essential degrees of freedom in magnets, e.g., charge, orbital, spin and lattice, the light has distinctive advantages to triggering magnetic phase transitions and even the emergence of hidden orders~\cite{Kirilyuk2010}. Besides, ultrafast optical control can realize magnetic phase transitions on subpicosecond time scales, providing a key technical support for the development of high-speed spintronic devices and ultrafast magnetic storage~\cite{Ghazaly2020}.

Pioneered by the ultrafast demagnetization of a Ni film~\cite{Beaurepaire1996}, various photoinduced nonequilibrium phenomena have been proposed and demonstrated in the past three decades~\cite{Meng2023,Wang2023}, such as light-induced coherent magnetization precession~\cite{Ju1999}, magnetic phase transitions~\cite{Agranat1998}, and hidden orders~\cite{Ichikawa2011}. 
In addition, light fields can also modulate other magnetism-related effects, including superconductivity~\cite{Jeon2018}, Floquet–Bloch states~\cite{Bielinski2025}, the Kondo effect~\cite{Wetli2018}, heavy-fermion states~\cite{Pal2019}, multiferroicity~\cite{Basini2024}, and magnetoresistance~\cite{Fiebig1998}.

Among the various physics, one of the most highly attractive and extensively studied families is double-exchange(DE) system, which is originally proposed by Zener, and later refined by Anderson and Hasegawa for ferromagnetic~(FM) oxides~\cite{Zener1951,Anderson1955}. Especially in colossal magnetoresistance materials, lots of novel experimental observations induced by light have been reported, including metal-insulator transition~\cite{Ogasawara2002}, time-scale separation between spin and charge~\cite{Matsubara2007}, charge and orbital order, lattice structural distortion~\cite{Beaud2009} and orthorhombicity evolution~\cite{Lu2022}, etc. 
Motivated by these findings, numerous physics unrealized in equilibrium have been proposed based on DE model and its extensions theoretically~\cite{Ishihara2019}.
Both continuous wave(cw) and pulse light fields can induce a FM to antiferromagnetic(AFM) phase transition on a square lattice, accompanied by narrowed electronic bands, a widened energy gap, and 1/4 filling electrons uniformly occupying the lower band\cite{Ono2017,Ono2018}. 
Cw light can also induce a 120$^\circ$ AFM phase transition on triangular lattice~\cite{Inoue2022}. 
Furthermore, nontrivial topological spin configurations have been proposed in various lattice geometries. An AFM chain can be driven to generate spin chirality by a typical pulse~\cite{Ghosh2022}. Photoinduced topological spin texture with a finite Pontryagin number and vector chirality can also be found in a square metallic ferromagnet~\cite{Ono2019}. 
S. Ghosh et al. excited the square antiferromagnet using a circularly polarized pulsed light field, generating meron-antimeron pairs that persist for over 100 ps~\cite{Ghosh2023}. Counterintuitively, when longer-range electronic hopping exists on a triangular lattice  centrosymmetrically, chirality-nontrivial spin configurations could be induced by a linearly polarized field.

However, most of these studies focus on a single photoinduced magnetic state. 
Thus, a question is raised: Can we modulate various electronic and magnetic states by varying the light field?
Previously, attentions have been paid to the strong DE interaction. Intuitively, in this coupling regime it is more feasible for electrons to drive local magnetic moments, due to their negligible coupling with light. However, a strong DE interaction and a FM initial equilibrium state typically result in a wide electronic band gap in equilibrium, which hinders inter-band electron transitions when irradiated by light. Furthermore, in order to acquire multiple photoinduced phases, we opt for the triangular lattice due to its relatively complex equilibrium phase diagram in the space of the electron filling and the strength of DE interaction~\cite{Azhar2017}.
Considering these two factors, we investigate the light-irradiated DE model in weak DE coupling regime on the triangular lattice. Only the impacts of amplitude, and frequency on the DE model are examined, while several novel electron and magnetic states have been uncovered.
The models and methods adopted are specifically presented in Section~\ref{model}. Our main results are shown in Section~\ref{result}. Section~\ref{sec:dis} provides some discussions and concludes this article, respectively.

\section{\label{model}model and methods}
\begin{figure}
    \centering
    \includegraphics[width=0.8\linewidth]{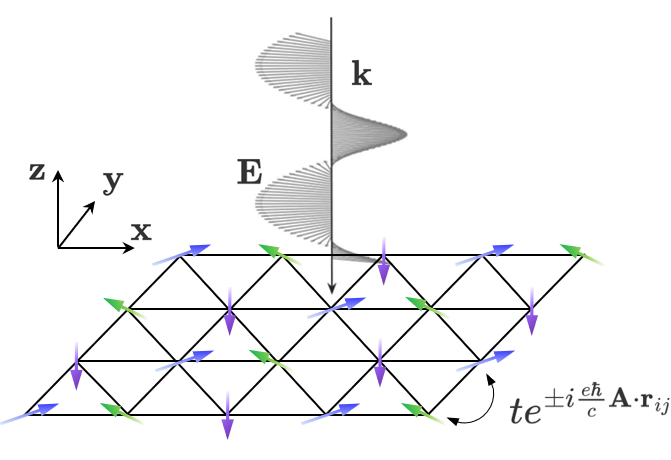}
    \caption{Cartoon of DE model irradiated by cw circular-polarized light. The blue, green, and violet arrows on triangular lattice denote the 120$^\circ$ coplanar groundstate in equilibrium. The light propagates along $\textbf{k}$, represented by the black arrow, while the gray ones show the electric field direction $\textbf{E}$ of circular polarized light at different times. Itinerant electrons are not shown, but their hopping integral $t$ on nearest-neighbor bond is displayed in the lower right corner.}
    \label{fig:cartoon}
\end{figure}

The Hamiltonian of light-irradiated DE model at time $\tau$ is written as
\begin{equation}
\begin{aligned}
    \mathcal{H}_{\tau} = & - \sum_{\langle ij\rangle,\alpha}(te^{i\frac{e\hbar}{c}\textbf{A}(\tau)\cdot\textbf{r}_{ij}}c_{i\alpha}^\dagger c_{j\alpha} +h.c.) \\
& + J\sum_{i\alpha\beta}c_{i\alpha}^\dagger \vec{\sigma}_{\alpha\beta} c_{i\beta}\cdot\vec{S}_i,
\end{aligned}
\end{equation}
where $t$ is the transfer integral between nearest neighbor site
$\langle ij\rangle$ in equilibrium, $e^{i\frac{e\hbar}{c}\textbf{A}(\tau)\cdot\textbf{r}_{ij}}$ comes from Peierls substitution when hopping electron couples with light. 
$\textbf{r}_{ij}=\textbf{r}_i-\textbf{r}_j$ labels the vector distance between site $i$ and $j$. 
$e,\ \hbar,\ c$ are physical constants. 
$\textbf{A}(\tau) = A_0\theta(\tau)[\sin(\omega\tau)\textbf{e}_x+\cos(\omega\tau)\textbf{e}_y]$, the vector potential of the circular polarized cw light, where $A_0,\ \omega$ are the amplitude and frequency of the electron field.
$c_{i\alpha}(c_{i\alpha}^\dagger)$ is annihilation(creation) operators of the itinerant electron on site $i$ with spin $\alpha$.
Vectors in spin space are indicated by arrows, while those in real space are denoted in boldface.
$J$ is the strength of DE interaction. $c^\dagger_{i\alpha}\vec{\sigma}_{\alpha\beta}c_{i\beta}$ and $\vec{S}_i$ denote the spins of itinerant electron and local magnetic moment at site $i$, where $\vec\sigma$ is Pauli spin vector and $\vec S_i$ is described by a unit vector in three dimensions. 
Fig.~\ref{fig:cartoon} shows the geometry at $\tau=0^-$, just before light reaches the lattice. 
The light propagates along $-\textbf{z}$, thus the electron field lies in lattice plane. 
$120^{\circ}$ coplanar AFM is chosen for equilibrium groundstate, corresponding to a relatively large electron density range in small $J$ regime~\cite{Azhar2017}.
Owing to computational considerations\cite{Ghosh2022}, we set the calculation temperature to $T\sim0.001$K; correspondingly, magnetic moments randomly deviate from the exact ground-state configuration by 0.1 rad~\cite{Koshibae2009,Ono2017}.

After the light arrives, the initial $\kappa$-th electron eigenstate $|\psi_\kappa(0)\rangle$ evolves as $|\psi_\kappa(\tau)\rangle=\exp(-i\cal{H}_{\tau}\tau)|\psi_\kappa($0$)\rangle$. 
Driven by electrons, magnetic moments move following Landau-Ginzburg-Gilbert Equation, $\partial_\tau \vec S_i=-J\langle\Psi(\tau)|\sum_{\alpha,\beta}c_{i\alpha}^\dagger\vec{\sigma}_{\alpha\beta}c_{i\beta}|\Psi(\tau)\rangle+\gamma\vec{S}_i\cdot\partial_\tau\vec{S}_i$, where $\gamma$ is a damping constant. 
The differential equation is solved by the forth order Runge-Kutta method. 
This electron and magnetic evolution method has been used in Ref.~\cite{Koshibae2009,Ono2017,Jeon2018,Ono2019,Ono2023}. 

$N_e$ electrons start with occupying the lowest $N_e/2$ bands, then the electron filling number in the $\nu$-th band at time $\tau$ is $\langle n_\nu\rangle=\sum_{\kappa}|\langle\phi_\nu(\tau)|\psi_\kappa(\tau)\rangle|^2$, where $|\phi_\nu(\tau)\rangle$ is the eigenstate of $\cal{H}_\tau$.
Besides, we calculate several order parameters to identify and characterize the magnetic states, including the magnetic structure factor $S(\textbf{q},\tau)=\frac{1}{N^2}\sum_{ij}\vec{S}_i(\tau)\cdot\vec{S}_j(\tau)e^{i\textbf{q}\cdot\textbf{r}_{ij}}$, scalar magnetic chirality $\chi_{ijkl}=\vec S_i\cdot(\vec S_j\times\vec S_k)+\vec S_j\cdot(\vec S_l\times\vec S_k)$ for a unit-cell plaquette with vertices $ijkl$, topological charge $Q=1/(2\pi)\sum_{ijk\in\Delta}\arctan(\vec S_i\cdot(\vec S_j\times\vec S_k)/(1+\vec{S}_i\cdot\vec{S}_j+\vec{S}_j\cdot\vec{S}_k+\vec{S}_k\cdot\vec{S}_i))$, and average angle between nearest-neighbor magnetic moments $\bar\theta=\frac{1}{6N}\sum_{i,\delta\in \rm{n.n.}}\arccos(\vec S_i\cdot\vec{S}_{i+\delta})$. For simplicity, evolution time index $\tau$ is omitted in the above formula.

The system is detected once every $\Delta\tau=0.02/t$, typically the measurements proceed up to $2000/t$. 
In regions of small $A_0$, where light-induced perturbation is weaker, we keep observing until the electron band and distribution stabilize. 
Here, the time unit $1/t$ is approximately 0.66fs for a typical value of transfer integral $t=1$eV and lattice constant $a=5\rm\AA$. 
Correspondingly, $t$ is the energy unit, and the electron field amplitude $E_0=A_0\omega$ can be expressed in the unit of $t/(ae)\approx20\rm MV/cm$\cite{Ono2023,Inoue2022}. 
$e \hbar/c$ is taken to be 1.
The total electron number $N_e=0.7N$ and DE interaction strength $J/t=1$ in our calculations, which ensure the equilibrium groundstate is the state shown in Fig.~\ref{fig:cartoon}\cite{Akagi2010,Azhar2017}. 
For lattice sizes $L=12$ and $L=18$ with periodic boundary conditions, our results are consistent with those reported in this paper. The damping constant $\gamma$ used in our calculations is $0.5/|\vec S_i|$, and magnetic moment length $|\vec{S}_i|=1$. 

\section{\label{result}results}
\begin{figure}
    \centering
    \includegraphics[width=0.8\linewidth]{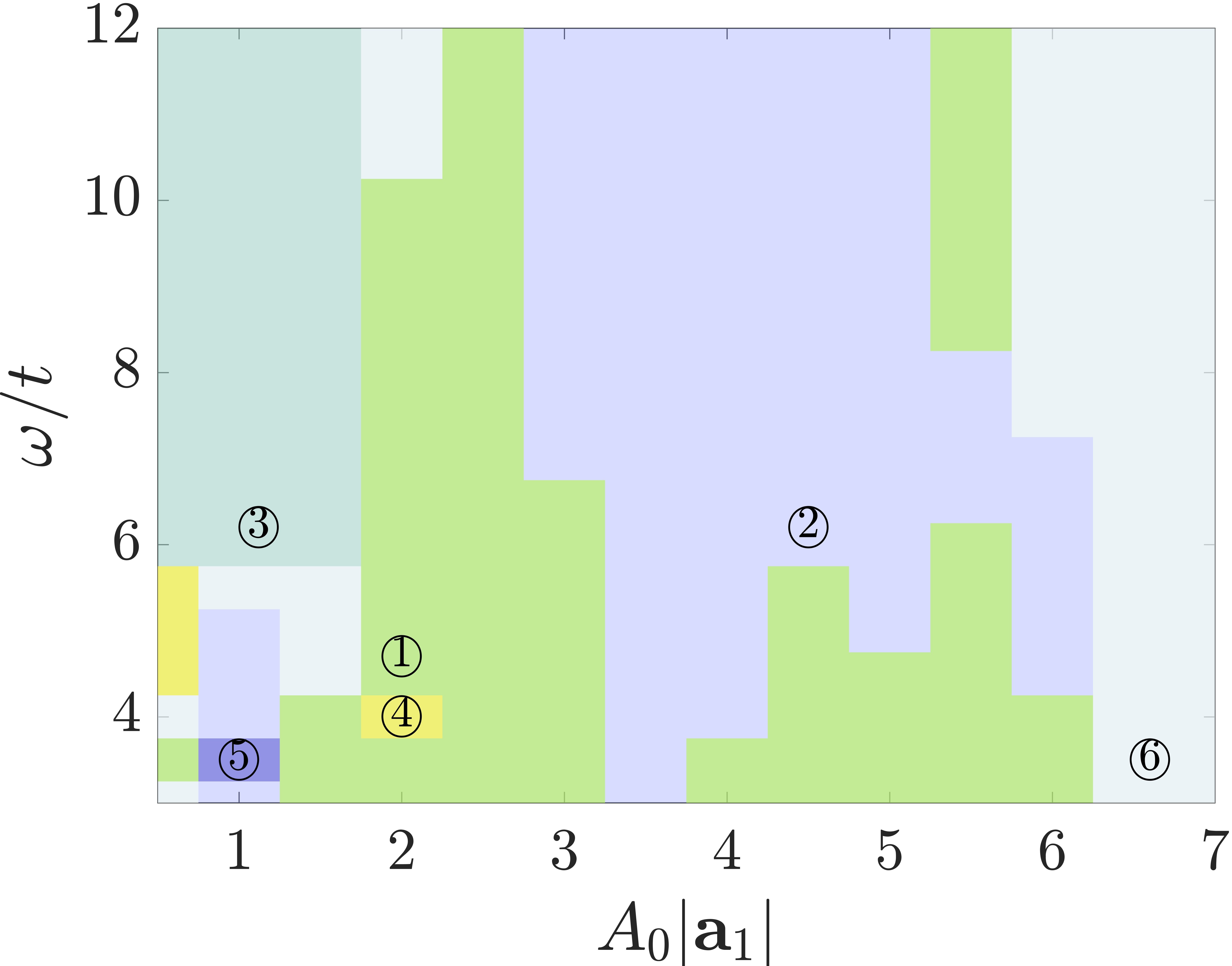}
    \caption{Phase diagram in $A_0$ and $\omega$ space. Different colors label various steady states. \textcircled{1}FM state. \textcircled{2}Vortex state. \textcircled{3}Quasi-long-range order at \textbf{K}. \textcircled{4}Long-Range Order at \textbf{K}/2 and $\textbf{K}^\prime/2$. \textcircled{5}Dynamical Long-Range Order at \textbf{M}. \textcircled{6}Disordered state.}
    \label{fig:phasediag}
\end{figure}
\subsection{Phase Diagram}
Six magnetic phases are found as we tune $A_0$ and $\omega$ in the range of $0.5\le A_0|\textbf{a}_1|\le7,\ 3\le\omega/t\le12$, where $\textbf{a}_{1,2}$ are basis vectors in real space.
Both calculation precisions are $0.5$ units. 
The phase diagram is shown in Fig.\ref{fig:phasediag}.
In the intermediate $A_0$ regimes, large‑scale photo‑induced FM and vortex states are found, labeled by $\textcircled{1}$ and $\textcircled{2}$, respectively. For larger $A_0$, the system is perturbed into a disordered state~$\textcircled{6}$.
While in small $A_0$ but large $\omega$ areas, a quasi-long-range order at $\textbf{K}$ with an "ordered" chirality, marked with $\textcircled{3}$, survives.
In the regions of small $A_0$ and $\omega$, several competing phases emerge, including long-range orders at $\textbf{K}/2~(\textbf{K}^\prime/2)$ \textcircled{4} as well as a dynamical long-range order at $\textbf{M}$, where both electron bands and magnetic moments vibrate with a special period, labeled by \textcircled{5}. 
The behaviors of electrons and magnetism in these states are looked over in following subsections.
\subsection{Ferromagnetic State}
\begin{figure}
    \centering
    \includegraphics[width=1.0\linewidth]{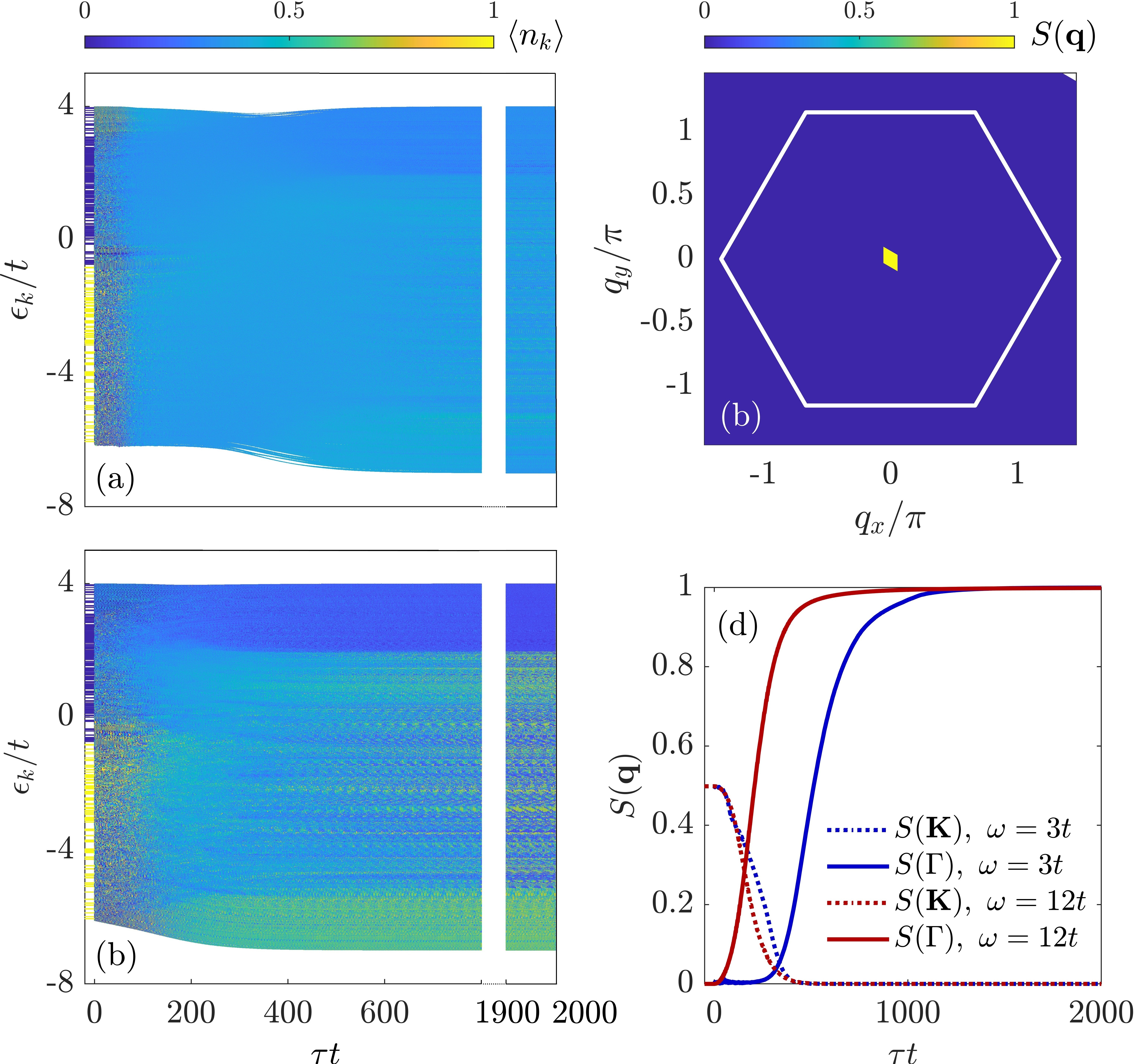}
    \caption{Behaviors of electrons and magnetic moments for $A_0|\textbf{a}_1|$=2.5, $\omega/t=3,12$ -- two cases of FM state in term of magnetism. (a)~Electron energy bands $\epsilon_k$ and fillings $\langle n_k\rangle$ for $\omega/t=3$. (b)~Magnetic structure factors $S(\textbf{q})$ peak at $\Gamma$ in the reciprocal space, the same for both cases. (c)~Electron energy bands for $\omega/t=12$. (d)~Magnetic structure factors $S(\textbf{K})$ and $S(\Gamma)$ vs. $\tau t$.}
    \label{fig:eg_FM}
\end{figure}
\begin{figure}[htbp]
\centering
\begin{overpic}[width=0.2\textwidth]{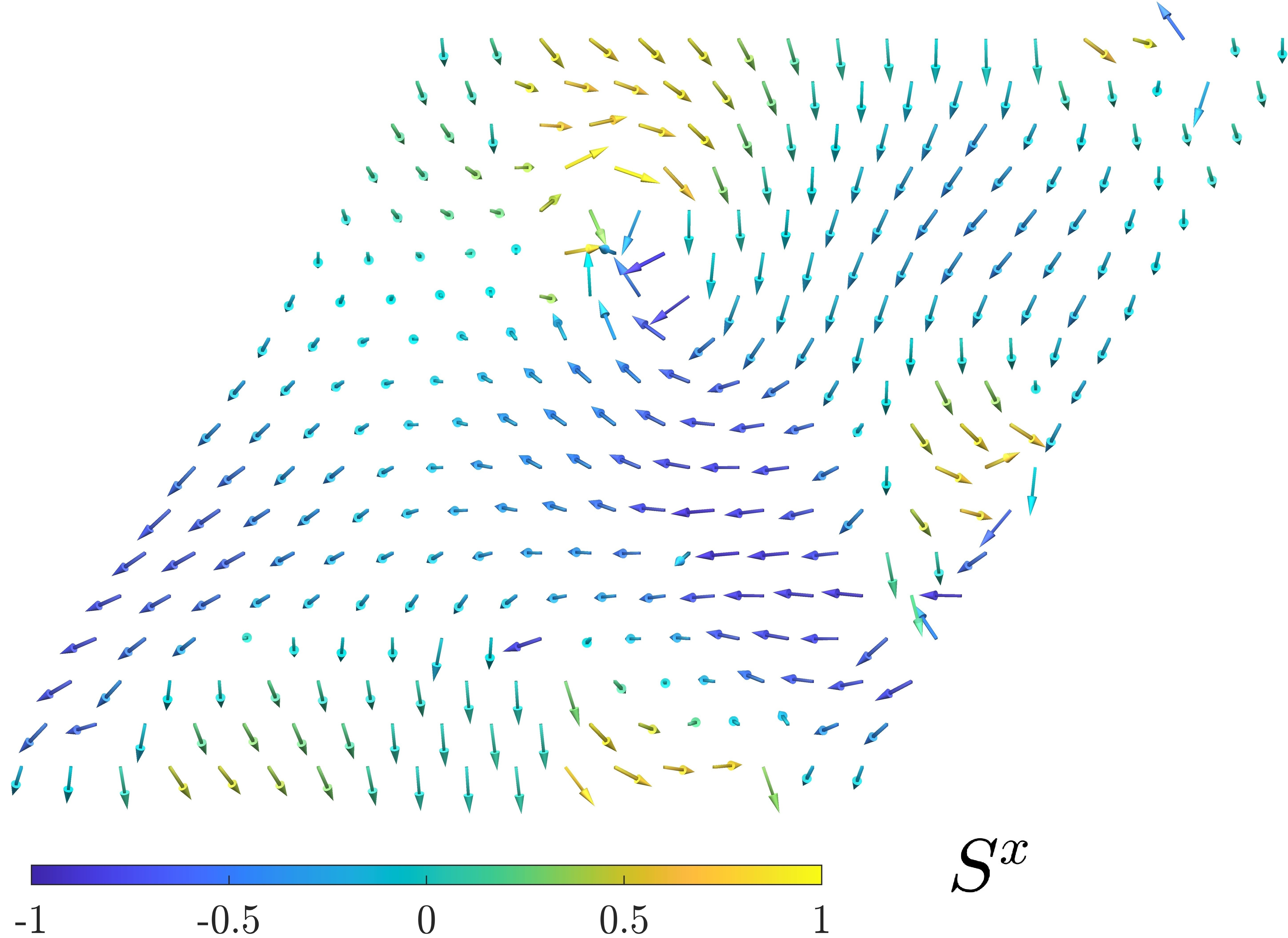}
\put(1, 65){(a)} 
\end{overpic}
\begin{overpic}[width=0.2\textwidth]{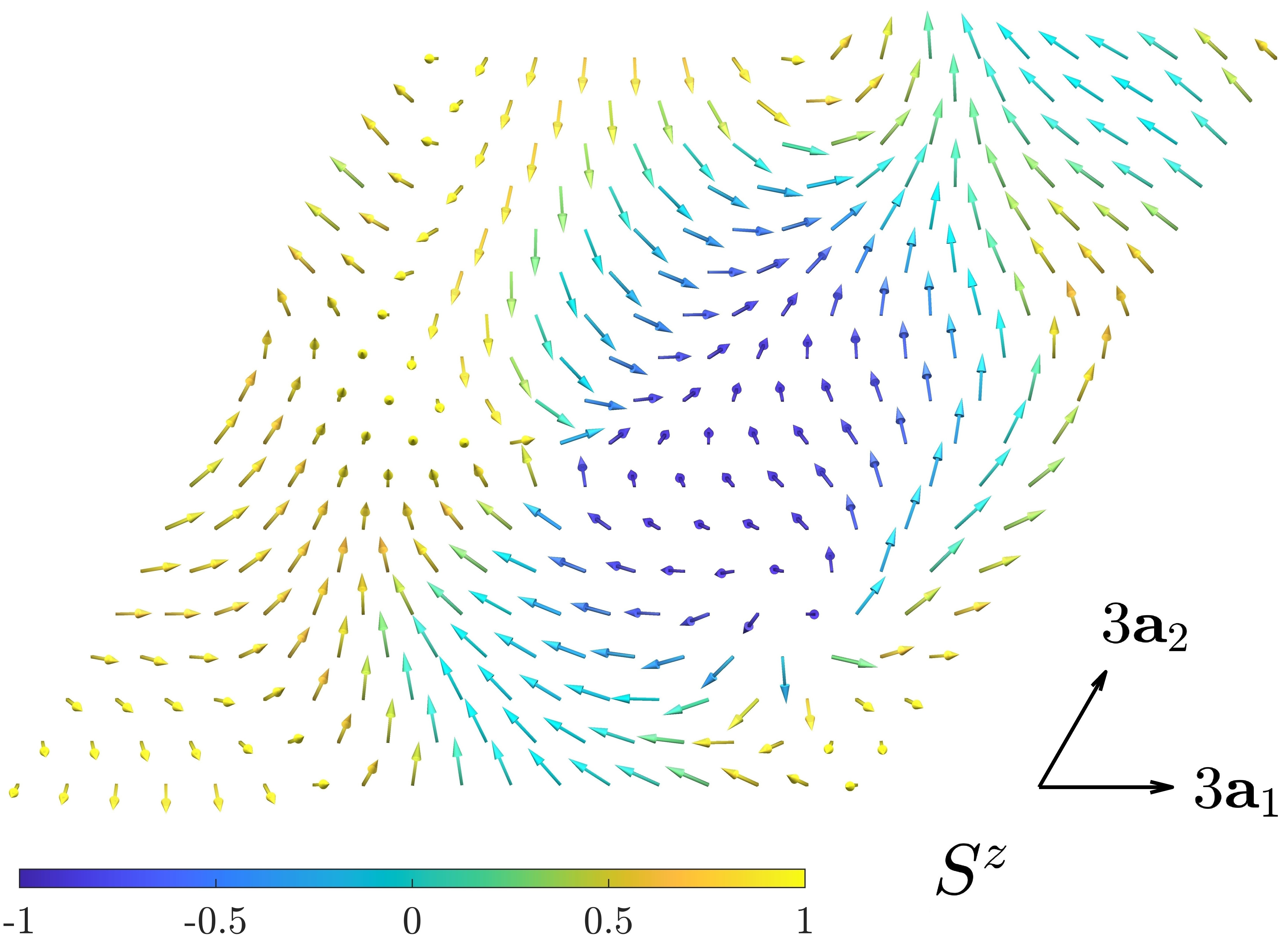}
\put(1, 65){(b)} 
\end{overpic}
\caption{Swirling structures for (a)~a intermediate state before FM forming with $A_0|\textbf{a}_1|=2.5,\omega/t=3$ and (b)~a stable vortex state with $A_0|\textbf{a}_1|=5.5,\omega/t=8.0$. The colorbars label $S^x$ in (a) and $S^z$ in (b). }\label{fig:spinconfig}
\end{figure}

FM order appears across almost the entire $A_0$ and $\omega$ ranges considered in our calculations. 
Two FM cases where $A_0|\textbf{a}_1|=2.5,\ \omega/t=3,12$ are shown in Fig.~\ref{fig:eg_FM}. 
Both $S(\textbf{q},\tau t=2000)$ peaks at $\Gamma$, where $S(\Gamma)>0.9$, as shown in Fig.~\ref{fig:eg_FM}(b). 
However, the electrons exhibit different behaviors in the two cases. 
Electrons are excited onto all energy bands almost evenly at weaker $\omega$~(Fig.\ref{fig:eg_FM}(a)), which resembles the system is heated up to a significant high temperature. 
As $\omega$ increases, electrons start to settle into the lower energy bands~(Fig.\ref{fig:eg_FM}(c)). 
Irritated by higher‑frequency light, the band broadens faster than it does for small $\omega$, though the light hardly influences the electrons’ highest energy for both cases. 
Besides, Fig.~\ref{fig:eg_FM}(d) reveals $S(\Gamma)$ grows more rapidly with increasing $\omega$, even though the equilibrium ground state whose $S(\textbf{K})$ dominants in $\textbf{q}$ space is suppressed as soon as the system is exposed to light for both cases.
By carefully examining the process of reaching stable FM order, we find that for small $\omega$, the magnetic configuration goes through a swirling state~(Fig.~\ref{fig:spinconfig}(a)), which is different from the case of large $\omega$ where FM interactions rapidly become dominant and directly align the local magnetic moments along the same direction.
Comparing (a) with (b) of Fig.~\ref{fig:eg_FM}, we find that energy band broadening implies a FM interaction mediated by electrons between magnetic moments, which is evidenced not only in Subsec.~\ref{sub:vortex} and \ref{sub:SK/2}, but also in other light-irradiated systems~\cite{Ono2017,Inoue2022}.

\subsection{Vortex State}\label{sub:vortex}
\begin{figure}
    \centering
    \includegraphics[width=1.0\linewidth]{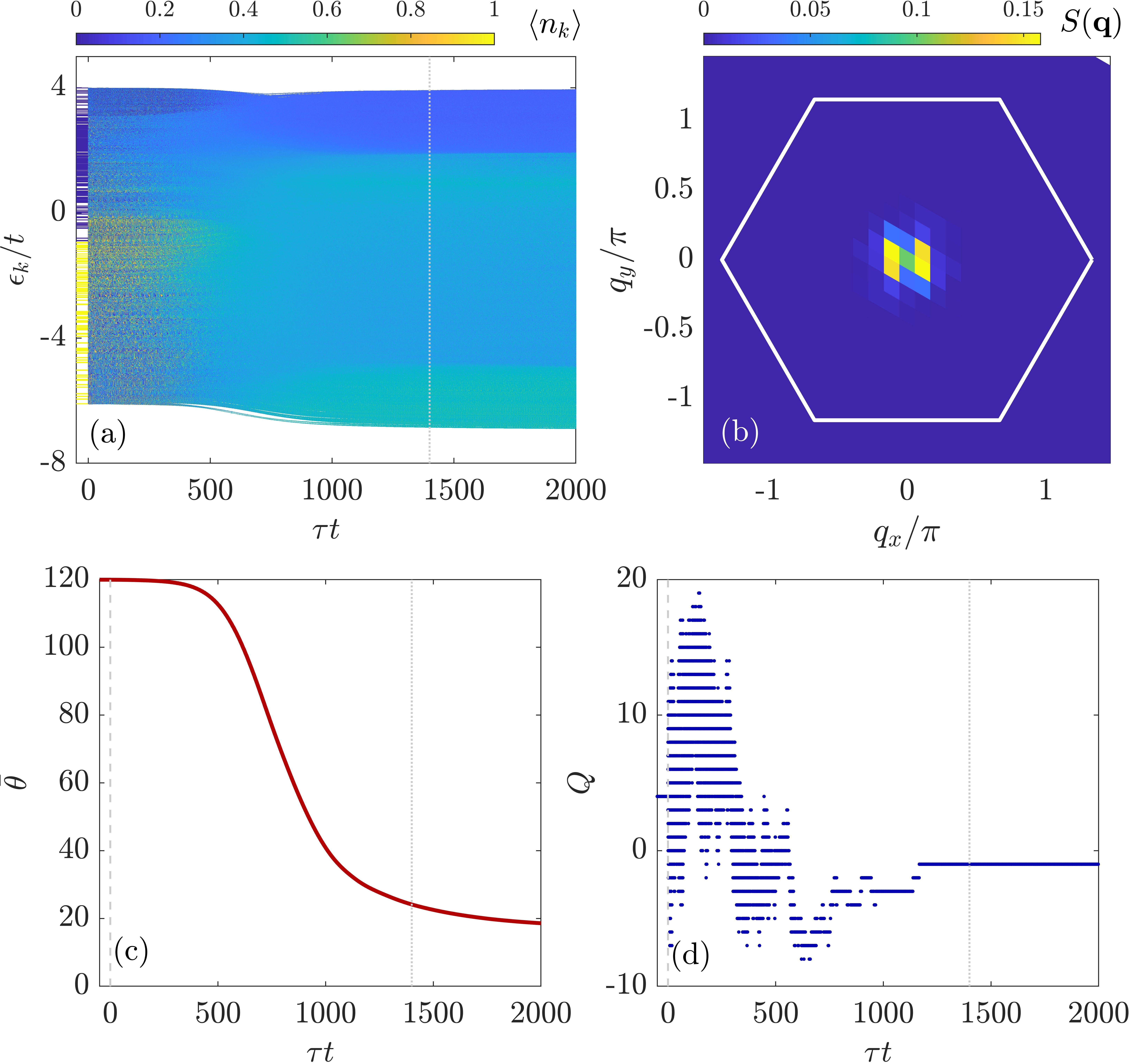}
    \caption{An example of vortex state with $A_0|\textbf{a}_1|=5.5,\ \omega/t=8.0$. (a)~Electron energy bands and fillings. (b)~Magnetic structure factors. (c)~The average angle between nearest‑neighbor magnetic moments $\bar\theta$ vs. $\tau t$. (d) Topological charge $Q$ vs. $\tau t$.}
    \label{fig:eg_vortex}
\end{figure}
Most vortex states occur in the moderate‑$A_0$ regime, in the vicinity of the FM state. 
As shown in Fig.~\ref{fig:eg_vortex}(a), after illuminating, the electrons are excited into all bands immediately. Subsequently, roughly in the range of $\tau t = 500$ to $1000$, the energy bands broaden and the electrons' distribution becomes layered. 
After approximately $\tau t = 1400$, the electronic states reestablish balance, thereby starting to stabilize the magnetic states. 
The energy bands here closely resemble that in the previous case. Nevertheless, FM interaction mediated by electrons persist but no longer prevails. Other interactions must compete with FM, driving the magnetic moments into a compromise state.
From another perspective, the vortex state is embedded within the FM component, reminiscent of the spiral state and skyrmion state in the 2D Kondo lattice model with Zeeman and anisotropy terms\cite{Wang2020}. This indicates that although the present light is sufficient to destroy the initial coplanar $120^\circ$ order at $\textbf{K}$ and generate multiple-$\textbf{q}$ states, the excited FM local magnetic moments effectively act as an external field, thus a vortex state exists\cite{Okubo2012,Leonov2015}.
Fig.~\ref{fig:eg_vortex}(b) records $S(\textbf{q})$ at $\tau t=2000$ for this steady state, where all relatively intense $S(\textbf{q})$ are concentrated on $\Gamma$ and its surrounding incommensurate momenta.
Direct snapshot of the magnetic moments reveals distinct vortex features, as illustrated in Fig.~\ref{fig:spinconfig}(b). 
We further investigate the average angle between nearest-neighbor magnetic moments $\bar\theta$ and the topological charge $Q$ of this state (Fig.\ref{fig:eg_vortex}(c–d)). We find that $\bar\theta$ stabilizes at $\sim20^\circ$, a significant reduction from $120^\circ$ in the equilibrium ground state. 
Meanwhile, $Q$ undergoes substantial fluctuations during light illumination before settling at $-1$. 
It has to be noted that within the parameter ranges considered herein, though most vortex configurations exhibit a total topological charge of $\pm1$ as we have shown above, some possess a charge of 0, and a negligible fraction of cases display $|Q|>1$. 

\subsection{Quasi-Long-Range Order at $\textbf{K}$}
\begin{figure}
    \centering
    \includegraphics[width=1.0\linewidth]{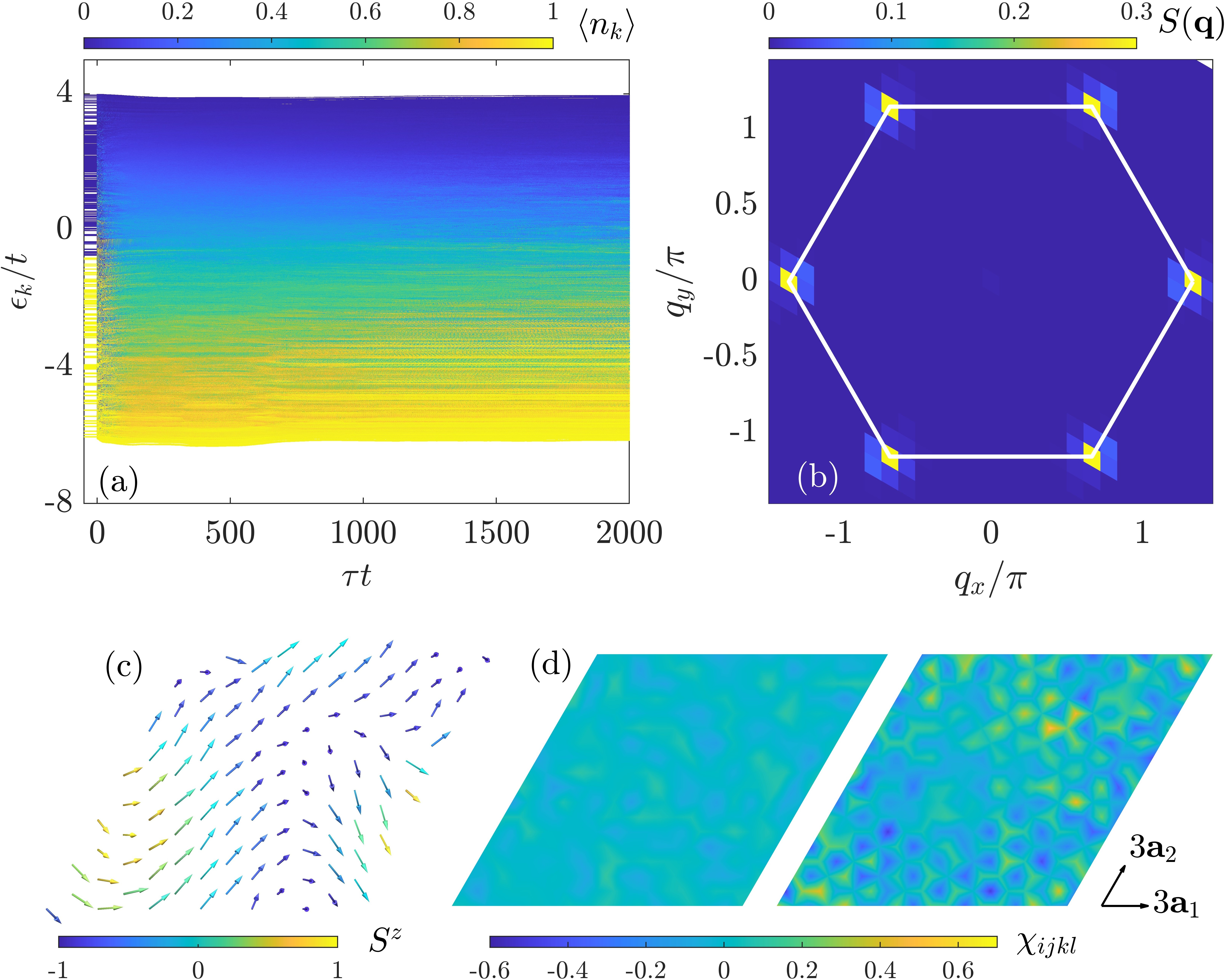}
    \caption{An example of quasi-long-range order at $\textbf{K}$ with $A_0|\textbf{a}_1|=1.0,\ \omega/t=11.0$. (a)~Electron energy bands and fillings. (b)~Magnetic structure factors. (c)~Magnetic configuration of one sublattice with topological swirling moment texture. (d)~The scalar magnetic chirality $\chi_{ijkl}$ at $\tau t=0$ and $2000$ is shown in the left and right panel, respectively.}
    \label{fig:eg_SK}
\end{figure}
As for the small $A_0$ and large $\omega$ areas, the energy bands first broaden slightly and then contract back to a width close to that of the equilibrium state. 
In contrast to other scenarios, the electron distribution behaves more like the Fermi distribution at a finite temperature with slight band changes-the higher the energy, the lower the electron fillings.
Concerning the magnetic configuration, $S(\textbf{q})$ is maximized at $\textbf{K}$ implying a $\sqrt{3}\times\sqrt{3}$ magnetic unit cell with 3 moments inside~(Fig.\ref{fig:eg_SK}(b)), just similar to the initial state. In addition, a ring of satellite peaks emerges around $\textbf{K}$, resembling the $S(\textbf{q})$ of vortex state in Fig.~\ref{fig:eg_vortex}(b). 
This quasi-long-range order at $\textbf{K}$ are consistent with the dynamical correlations at finite energy of the triangular Heisenberg antiferromagnet, which is a signature of the Z$_2$-vortex excitation\cite{Okubo2010}.
Fig.\ref{fig:eg_SK}(c) offers the magnetic moments of one of the three sublattices at $\tau t = 2000$ and as expected, where topological swirling moment texture is exhibited.
Thus, we argue due to the small $A_0$, the excited electrons can not destroy completely destroy the equilibrium ground state, but elevate moments to higher energy, which manifests as the introduction of other interactions competing with the AFM one.
The topological swirling state appears here and Subsec.\ref{sub:vortex}—at least within the range of our investigations—is stable, which is different from the situation in Subsec.\ref{fig:eg_FM} and Ref.~\cite{Ono2019}. 
Beyond the swirling structure formed by the same sublattices, the scalar chirality $\chi_{ijkl}$ in one magnetic unit cell is enhanced , which indicates the magnetic moments are no longer coplanar despite there is still $\bar{\theta}\sim 120^\circ$~(not shown in figure). Besides, while each individual sublattice adopts a swirling texture, the $\chi_{ijkl}$ displays emergent long-range order, which implies a periodic distribution of solid angles formed by every three neighboring moments, as shown in the right panel in Fig.\ref{fig:eg_SK}(d). 
\subsection{Long-Range Order at \textbf{K}/2 and $\textbf{K}^\prime/2$}\label{sub:SK/2}
\begin{figure}
    \centering
    \includegraphics[width=1.0\linewidth]{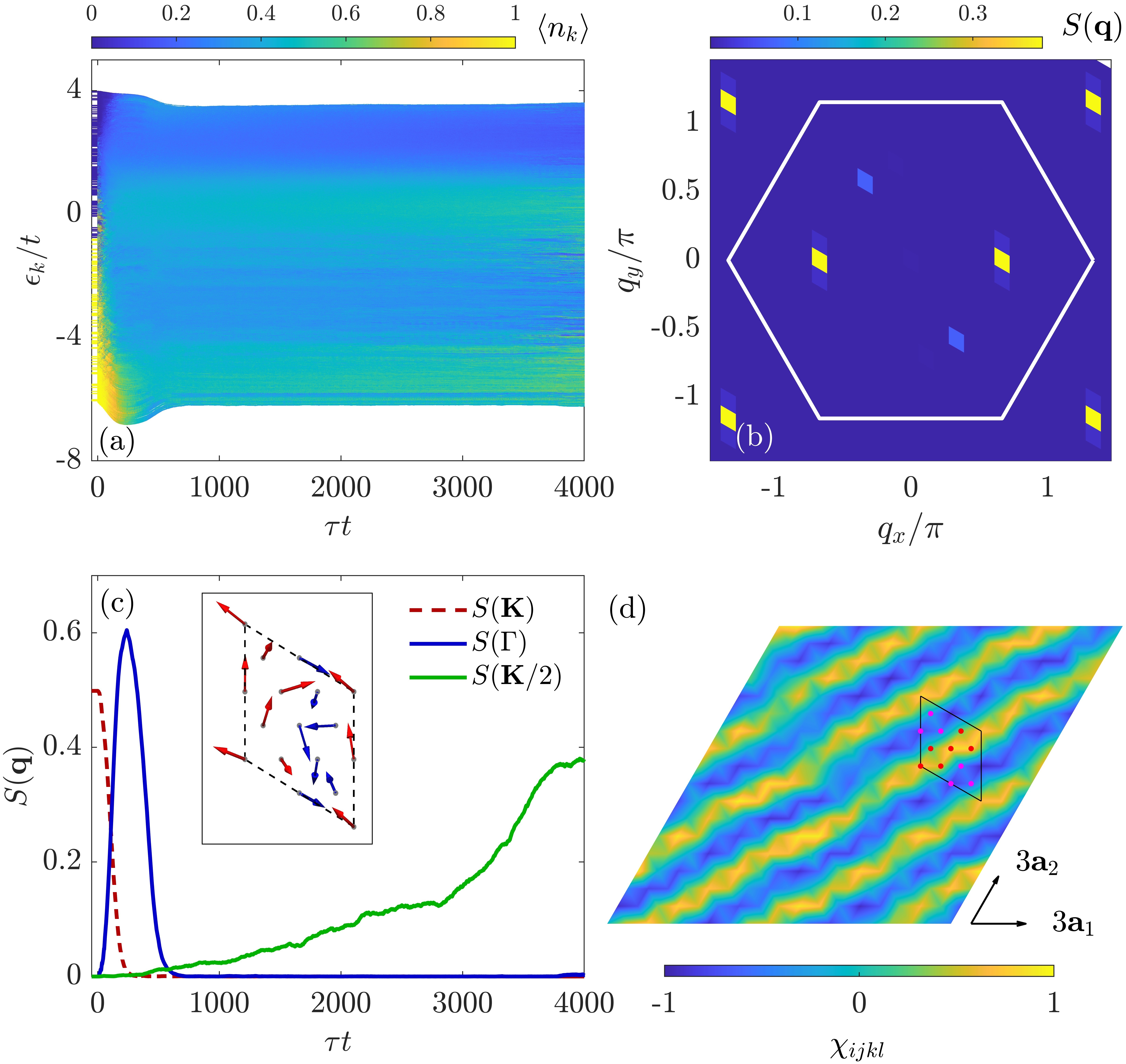}
    \caption{An example of long-range order at \textbf{K}/2 and $\textbf{K}^\prime/2$ with $A_0|\textbf{a}_1|=0.5,\ \omega/t=4.5$. (a)~Electron energy bands and fillings. (b)~Magnetic structure factors. (c)~Magnetic structure factors $S(\textbf{K}),\ S(\Gamma),\ S(\textbf{\textbf{K}/2})$ vs. $\tau t$. A magnetic unit cell plotted by black dashed line is inserted. Magnetic moments representing by red and blue arrows with $(3\textbf{a}_1,-\sqrt{3}\textbf{a}_2)/2$ shifted are opposite. (d)~Scalar Magnetic chirality of upright triangles. The chirality of inverted triangles has the same pattern and is not shown here. One of the magnetic unit cells is indicated by a solid black line, and the points inside mark the positions of the triangular lattice sites. The red and magenta markers denote the upright triangles with opposite signs of $\chi_{ijk}$.}
    \label{fig:eg_SK/2}
\end{figure}
This long-range order is primarily characterized by maxima in $S(\textbf{q})$ at $\textbf{K}/2$ and $\textbf{K}^\prime/2$ (Fig.~\ref{fig:eg_SK/2}(b)), corresponding to a $2\sqrt{3}\times2\sqrt{3}$ magnetic unit cell containing 12 sublattices in real space~(drawn in the inset of Fig.~\ref{fig:eg_SK/2}(c)). 
Upon photoexcitation, electrons rapidly populate higher bands, and the band width first increases sharply before narrowing back to a value close to the equilibrium one by $\tau t=700$ (Fig.~\ref{fig:eg_SK/2}(a)). 
The fillings of bands stabilize into four layers. 
Unlike earlier cases, all bands are shifted downward, especially for the highest level. 
Consistent with our previous observations, band broadening signals a FM tendency, which is corroborated by the behavior in Fig.~\ref{fig:eg_SK/2}(c) as well. 
As soon as the light arrives, $S(\Gamma)$ is strongly enhanced while $S(\textbf{K})$ is immediately suppressed, peaks when $S(\textbf{K})$ vanishes, then decays rapidly. 
Concurrently, $S(\textbf{K}/2)$ grows slowly, eventually dominates, and establishes stable long-range order. 
The basis vectors of the magnetic unit cell are $2\textbf{a}_1-\textbf{a}_2$ and $2\textbf{a}_2-\textbf{a}_1$, as shown in Fig.~\ref{fig:eg_SK/2}(d). 
Within a unit cell, the moments display pronounced AFM correlations along $2\textbf{a}_1-\textbf{a}_2$, whereas such correlations are negligible small along the other direction, in agreement with the reciprocal-space peaks in Fig.~\ref{fig:eg_SK/2}(b). 
This configuration induces a periodic modulation of $\chi_{ijk}$ along $\textbf{a}_2-\textbf{a}_1$, shown in Fig.~\ref{fig:eg_SK/2}(d), where $\chi_{ijk}=\vec S_i\cdot(\vec S_j\times\vec S_k)$, calculated by magnetic moments on upright triangles with vertices $i,j,k$. 
Although no clear AFM order is observed along $2\textbf{a}_2-\textbf{a}_1$, the chiralities $\chi_{ijk}$ of the moments are opposite between two neighboring triangles along both basis directions, as inferred from $\chi_{ijk}$ and lattice geometry in Fig.~\ref{fig:eg_SK/2}(d). To the best of our knowledge, such a long-range order has not yet been reported on the triangular lattice.

\subsection{Dynamical Long-Range Order at $\textbf{M}$}
\begin{figure}
    \centering
    \includegraphics[width=1.0\linewidth]{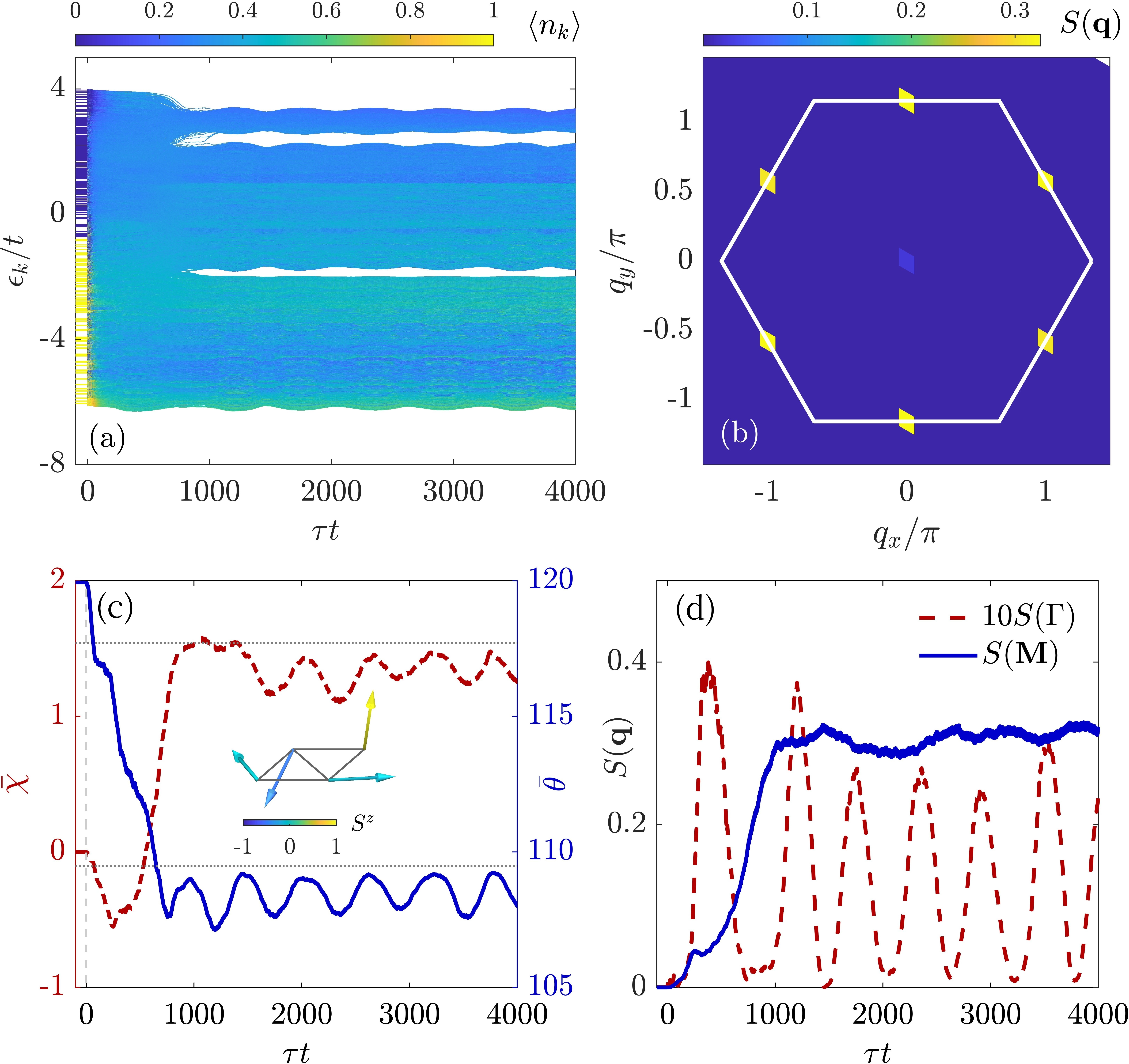}
    \caption{An example of dynamical long-range order at $\textbf{M}$ with $A_0|\textbf{a}_1|=1.0,\ \omega/t=3.5$. (a)~Electron energy bands and fillings. (b)~Magnetic structure factors. (c)~Average scalar magnetic chirality $\bar\chi$ vs. $\tau t$~(red dashed line) and the average angle of nearest‑neighbor magnetic moments $\bar\theta$ vs. $\tau t$~(blue solid line). The gray dotted lines are calculated by magnetic moments pointing to vertices of the regular tetrahedron. The inserted is the magnetic configuration in a unit cell. (d)~Magnetic structure factors $S(\textbf{M})$ and $10\times S(\Gamma)$ vs. $\tau t$.}
    \label{fig:eg_SM}
\end{figure}

A long-range ordered state at \textbf{M}, occupying a very small area in the phase diagram, exhibits periodic oscillations. 
To assess its stability, we examine the state up to $\tau t=4000$. 
As shown in Fig.~\ref{fig:eg_SM}(a), two energy gaps open near $\pm 2t$ after the system enters this state. 
The electron density is the highest in the lowest band, intermediate in the middle band, and the upper band is the narrowest with the lowest electron density. But such an electron distributions across the three bands are relatively uniform comparing the ones of larger $\omega$ excitations in other cases.
The band-edge dynamics further indicate an oscillation period much longer than the optical frequency. 
Fig.~\ref{fig:eg_SM}(b) displays a clear long-range order at the $\textbf{M}$ and only very weak order at the $\Gamma$. 
Despite the oscillatory behavior of the electronic structure, $S(\textbf{M})$ does not exhibit periodic oscillations during illumination (Fig.~\ref{fig:eg_SM} (d)), implying that the spatial periodicity of the 4-sublattice state remains stable~(shown in the inset of Fig.~\ref{fig:eg_SM}), which resembles the scalar chiral state reported in Ref.~\cite{Ono2023} and exhibits an anomalous Hall effect~\cite{Akagi2010}.
Thus dynamical interactions between moments mediated by oscillating electrons modifies the magnetic configuration within a magnetic unit cell, manifesting as a weak oscillating FM order when the 4 moments approach and depart from the regular-tetrahedron limit, which is demonstrated by $\bar\chi$ and $\bar\theta$ in Fig.~\ref{fig:eg_SM}(c). 

From the phase diagram (Fig.~\ref{fig:phasediag}), we find that within the scope of our study, the system excited by circularly polarized light does not favor a stable long-range order at $\textbf{M}$. 
Nevertheless, we will demonstrate in a subsequent work that via optical modulation, a stable long-range order at $\textbf{M}$ and the quasi-long-range order with nontrivial topological features can still be realized on this triangular metallic AFM magnet.

\section{\label{sec:dis}Discussions and Conclusions}
We have uncovered various novel magnetic orders in the simple photoexcited triangular metallic magnet, including the FM state, vortex state, quasi-long-range order at $\textbf{K}$, long-range order at $\textbf{K}/2$ and $\textbf{K}^\prime/2$, and dynamical long-range order at $\textbf{M}$. 
We then provide a detailed analysis of the corresponding electronic characteristics and magnetic configurations.
In contrast to previous studies, we argue that a weaker DE interaction and a AFM equilibrium state give smaller energy band gap. Thus it is more favorable for the optical field to induce electronic excitations. 
Meanwhile, the abundant equilibrium magnetic states in the phase diagram of $J$ and $n$ space on triangular DM model also indicate that the magnetic configuration is extremely sensitive to the electronic state, which facilitates the observation of these rich electronic and magnetic phases~\cite{Akagi2010}.

Low-frequency excitations in the range $0<\omega<3$ are investigated as well, but not presented in this paper. In $A_0-\omega$ phase diagram of this low-$\omega$ region, weak FM orders are dominant, with disordered states scattered throughout.

To realize all these states, we estimate the requisite electric field strength $E_0=A_0\omega$ is on the order of $1\sim100$~MV/cm. 
However, the continuous-wave optical intensity achievable in the laboratory currently is just on the order of 10~MV/cm. 
Besides, field intensities of this magnitude are sufficiently high to induce material degradation. This issue can probably be addressed in two ways. 
For one thing, periodic terahertz pulse can be employed instead of continuous-wave light in the hope of achieving a similar effect~\cite{Fulop2020}. 
For another, other $C_6$-invariant lattices with more fragile magnetic structures can be chosen. Both of these aspects have been explored preliminarily in our subsequent work.

\section{\label{ack}Acknowledgement}
I thank Yuan Wan for useful discussions. This work was interrupted by unexpected health complications. During this period, I must express my sincere gratitude to my parents for their love, care, and unwavering support. This work is supported by the National Science Foundation of China Young Scientists Fund (Grant No. 12504179) and Shandong Provincial Natural Science Foundation Youth General Special Project (Grant No. ZR2025QC1500).

\bibliography{apssamp}

\end{document}